# A Methane Extension to the Classical Habitable Zone

Ramses M. Ramirez[1,2], and Lisa Kaltenegger[2,3]

[1]Earth-Life Science Institute, Tokyo Institute of Technology, Tokyo, Japan
[2]Carl Sagan Institute, Cornell University, Ithaca, NY
[3]Department of Astronomy, Cornell University, Ithaca, NY

Corresponding author - Ramses Ramirez email: rramirez@elsi.jp

**ABSTRACT**

The habitable zone (HZ) is the circumstellar region where standing bodies of liquid water could exist on the surface of a rocky planet. Conventional definitions assume that $CO_2$ and $H_2O$ are the only greenhouse gases. The outer edge of this classical $N_2$-$CO_2$-$H_2O$ HZ extends out to nearly ~1.7 AU in our solar system, beyond which condensation and scattering by $CO_2$ outstrip its greenhouse capacity. We use a single column radiative-convective climate model to assess the greenhouse effect of $CH_4$ (10 - ~100,000 ppm) on the classical HZ ($N_2$-$CO_2$-$H_2O$) for main-sequence stars with stellar temperatures between 2,600 – 10,000 K (~A3 to M8). Assuming $N_2$-$CO_2$-$H_2O$ atmospheres, previous studies have shown that cooler stars more effectively heat terrestrial planets. However, we find that the addition of $CH_4$ produces net greenhouse warming (tens of degrees) in planets orbiting stars hotter than a mid-K (~ 4500K), whereas a prominent anti-greenhouse effect is noted for planets around cooler stars. We show that 10% $CH_4$ can increase the outer edge distance of the hottest stars ($T_{EFF}$ =10,000 K) by over 20%. In contrast, the $CH_4$ anti-greenhouse can shrink the HZ for the coolest stars ($T_{EFF}$ = 2,600 K) by a similar percentage. We find that dense $CO_2$–$CH_4$ atmospheres near the outer edge of hotter stars may suggest inhabitance, highlighting the importance of including secondary greenhouse gases in alternative definitions of the HZ.

## 1. INTRODUCTION

The habitable zone (HZ) is the circular region around one or multiple stars where standing bodies of liquid water could exist on a rocky planet's surface (e.g. Kasting et al., 1993, Kaltenegger & Haghighipour 2013, Haghighipour & Kaltenegger 2013) and facilitate the detection of possible atmospheric biosignatures (see e.g. Kaltenegger 2017).[1] Unlike earlier studies (e.g. Hart, 1978) which computed a relatively narrow habitable zone, Kasting et al. (1993) showed that the carbonate-silicate cycle, which regulates the transfer of $CO_2$ between the surface, atmosphere, and interior on the Earth, is what allows for a relatively wide (~0.95 – 1.67 AU in our solar system) region for the classical HZ, which is calculated using a single column radiative-convective model.

The classical $N_2$-$CO_2$-$H_2O$ habitable zone is defined by the greenhouse effect of two gases: $CO_2$ and $H_2O$ vapor. The inner edge corresponds to the distance where mean surface temperatures exceed the critical point of water (~647 K, 220 bar), triggering a runaway greenhouse that leads to rapid water loss to space on very short timescales (see Kasting et al.,1993 for details). Towards the outer edge of the classical habitable zone, weathering rates decrease, allowing atmospheric $CO_2$ concentrations to increase as stellar insolation decreases. At the outer edge, condensation and scattering by $CO_2$ outstrips its greenhouse capacity, the so-called maximum greenhouse limit of $CO_2$.

The inner edge of the classical habitable zone appears to be relatively robust to additions of secondary greenhouse gases because trace gas absorption is outstripped by water vapor absorption in these extremely dense (> 200 bar) water steam atmospheres (Ramirez & Kaltenegger 2017). In contrast, the outer edge limit of the $N_2$-$CO_2$-$H_2O$ habitable zone can change significantly if other gases are added to the model. For example, if volcanic hydrogen is outgassed in sufficiently high amounts on a planet, it can increase the orbital distance of the outer edge distance of the classical HZ by well over 50 %. In addition, adding a light gas like hydrogen to the atmosphere would improve detectability of atmospheric features on outer edge planets (Ramirez and Kaltenegger, 2017), making super-Earths beyond the traditional HZ compelling observational targets as well. The accumulation of tens to hundreds of bars of primordial hydrogen in the atmosphere of a young planet can increase this distance even farther, however the duration of habitable conditions within the classical HZ in this case is very short (see Pierrehumbert and Gaidos 2011).

Many studies have shown that the stellar energy distribution (SED) of a star (Fig.1) influences both the magnitude and location of atmospheric warming (e.g., Kasting et al., 1993; Kopparapu et al. 2013; von Paris et al. 2010; Wordsworth et al. 2011; Hu and Ding, 2011; Ramirez and Kaltenegger, 2014; Ramirez and Kaltenegger, 2016). In comparison to terrestrial planets around G stars, those around M stars absorb more stellar energy, whereas those around A and F stars absorb less. This is partly due to the effectiveness of Rayleigh scattering, which decreases at longer wavelengths. A second effect is the increase in near-IR absorption by $H_2O$ and $CO_2$ as the star's spectral peak shifts to these wavelengths. That means that the same integrated stellar flux that hits the top of a planet's atmosphere from a cool red star warms a planet more efficiently than the same integrated flux from a hot blue star. However, these previous studies have only analyzed atmospheres consisting of $N_2$, $CO_2$ and $H_2O$, with neither gas possessing near-infrared absorption features that are similar in strength to their dominant IR features in the temperature regime typically considered for habitable planets (~ 150 – 300 K). In contrast, methane has several absorption bands in the near-IR that are comparable in strength to its IR absorption bands as well as a scattering cross section 2.4 times that of air (Sutton et al. 2004; Pavlov et al. 2000) (see Fig.2).

---

[1] In spite of its tenuous atmosphere (~ 6mb), Mars' surface can support seasonal trickles of surface water (Stillman et al., 2016). However, sufficiently large standing bodies of liquid water are needed for biosignature detection.

Plus, $CH_4$ has been invoked to solve the faint young sun problem on the early Earth (e.g. Pavlov et al. 2000; Haqq-Misra et al.; 2008; Wolf and Toon, 2013). Methane concentrations high enough to generate above-freezing surface temperatures in these $N_2$-$CO_2$-$H_2O$-$CH_4$ atmospheres may have been produced by methanogens. Likewise, it has been proposed that $CH_4$ could have been a key greenhouse gas on early Mars (e.g. Wordsworth et al., 2017; Ramirez, 2017). As a trace gas on the modern Earth, $CH_4$ is notable because ~90% of the present-day atmospheric $CH_4$ on our own planet has a biotic origin.

In this paper, we explore the effect of $CH_4$ on the habitability of terrestrial planets and its effect of the limits of the classical $N_2$-$CO_2$-$H_2O$ habitable zone. We model the effects of $CH_4$ on the temperature structure of various terrestrial atmospheres as a function of SED with a single column radiative-convective climate model. We then calculate the $N_2$-$CO_2$-$H_2O$-$CH_4$ HZ limits for A – M-stars ($T_{eff}$ = 2,600 – 10,000 K) and parameterize these limits. Finally, we discuss the implications of life near the outer edge and its relevance to atmospheric detection and characterization. Our models are described in section 2, section 3 describes our results, and section 4 discusses the implications of this work. We then summarize with concluding thoughts.

## 2. METHODOLOGY

As in previous studies of the classical HZ (e.g. Kopparapu et al., 2013; Ramirez and Kaltenegger, 2014; 2016; 2017) we used a single-column radiative-convective climate model to compute habitable zone boundaries for stars of stellar effective temperatures, $T_{EFF}$, ranging from 2,600K to 10,000K (Ramirez and Kaltenegger, 2016). The HZ limits we compute here (and in Ramirez and Kaltenegger, 2016; 2017) include A-stars, with larger orbital separations of the HZ. We use spectra derived from BT-Settl data (Allard et al., 2003; 2007) and a standard Thekaekara solar spectrum for the sun-like (G2) star (Thekaekara, 1974).

The model uses correlated-k coefficients to parameterize absorption by $H_2O$, $CO_2$, and $CH_4$ across 38 spectral intervals at solar wavelengths and 55 intervals in the infrared. The near-IR $CH_4$ coefficients for wavelengths less than 1 micron are derived from Karkoschka (1994). As in Ramirez (2017), we use HITRAN 4-term $CH_4$ infrared absorption coefficients derived for 8 pressures ($1\times10^{-5}$- 100 bar) and 5 temperatures (100, 200, 300, 400, and 600 K). We include new $CO_2$-$CH_4$ collision induced absorption data (Wordsworth et al. 2017), which has been recently shown to greatly increase the efficacy of the $CH_4$ greenhouse (Wordsworth et al. 2017; Ramirez, 2017). Although we also included the most updated $CH_4$-$CH_4$, $N_2$-$CH_4$, and $N_2$-$N_2$ CIA (Richard et al., 2012), previous sensitivity studies showed that the additional forcing from these CIA is negligible for these atmospheres (Ramirez, 2017). We also added

Rayleigh scattering due to $CH_4$ based on experimentally-derived indices of refraction (Sneep & Ubachs 2005).

In previous 1-D radiative-convective climate modeling of the outer edge of the habitable zone, inverse calculations (e.g. Kasting et al., 1993) were typically used, in which a surface temperature is specified and the solar flux required to maintain it is computed. In those earlier computations, the surface and stratospheric temperatures were both fixed at 273K and ~155 K, respectively (Kasting et al., 1993; Kopparapu et al., 2013) and the $CO_2$ partial pressure was varied from $1\times10^{-2}$ to 34.7 bar (the saturation $CO_2$ partial pressure at 273K). The effective fluxes incident on the planet ($S_{EFF}$) to sustain a surface temperature of 273 K are then computed for all spectral classes. For the inner edge, the stratospheric temperature is 200 K and the surface temperature is gradually increased from a starting temperature of 200 K, which simulates pushing the planet closer and closer to the star until a runaway greenhouse is triggered, following Kasting et al. (1993) and Ramirez and Kaltenegger (2017). An Earth-like $CO_2$ concentration of 330 ppm is assumed. A background of 1 bar $N_2$ is assumed for all atmospheres (ibid).

However, assuming a constant stratospheric temperature profile for $CH_4$ is inaccurate because significant absorption at solar wavelengths produces upper atmospheric temperature inversions (see Results). We instead employ the method of forward calculations for these $CO_2$-$CH_4$-$H_2O$ atmospheres (e.g. Kasting, 1991). With this method, we specify a solar flux and compute the surface temperature once stratospheric fluxes are balanced and convergence is achieved.

We pick several representative stars from the A-M spectral classes (2,600K, 3,400K, 3,800K, 4,400K, 5,000K, 5,800K, 7,200K, and 10,000 K) and calculate the effective flux needed to converge to a mean surface temperature of 273 K, the freezing point of water, due to a gradual increase in $pCO_2$. For $pCO_2 < 1$ bar, we assume the following grid of atmospheric $pCO_2$ levels (in bar) [$1\times10^{-3}$, $5\times10^{-3}$, $1\times10^{-2}$, $5\times10^{-2}$, $1\times10^{-1}$, $5\times10^{-1}$]. At higher $pCO_2$ levels, pressure is gradually increased by whole units. This is a much more time-intensive procedure than inverse calculations because a convergent solution must be obtained for each $pCO_2$ level before the solution for the next one can be computed. If a mean surface temperature of 273 K is not obtained on a given initial stellar flux, the calculation must be repeated using a different estimate on the stellar flux for the entire suite of $pCO_2$ levels.

We compute the resultant outer edge effective stellar incident flux limits of the $N_2$-$CO_2$-$H_2O$-$CH_4$ HZ for A3 – M8 stars ($T_{eff}$ = 2,600 – 10,000 K) and for 3 $CH_4$ concentrations: 10 ppm, 10,000 ppm, and an upper limit equal to 10% of the $CO_2$ abundance, above which photochemical hazes should form that cool the planet (Haqq-Misra et al. 2008). Although recent results suggest that haze formation may be inhibited on HZ planets orbiting stars with high FUV fluxes (A and F spectral classes and active M-stars) (Arney et al. 2017), this result will need to be verified with a more complex photochemical model that incorporates oxygen into the haze molecules (ibid). We nominally assume that hazes can form around such stars for this analysis. The 10,000 ppm (1% $CH_4$) case is a limit for Earth-like planets assuming terrestrial outgassing rates and diffusion-limited escape (Pavlov et al., 2000; Kharecha et al. 2005). For the inner edge, even the highest $CH_4$ concentration yields 33 ppm (~ 1/300 of a percent $CH_4$) for the dry ($N_2$ + $CO_2$) atmosphere. We verified that at such low $CH_4$ concentrations this inner edge calculation would be virtually indistinguishable from that predicted with the classical $N_2$-$CO_2$-$H_2O$ HZ. Thus, we focus our work on the outer edge.

As explained in Kasting et al. (1993), the effective flux $S_{EFF}$, is the ratio of the net outgoing infrared radiation ($F_{ir}$) over the net incoming solar radiation ($F_s$). An $S_{EFF}$ value of 1 is the normalized flux received at Earth's orbit. Higher $S_{EFF}$ values correspond to distances closer to the star. Given $S_{EFF}$, the stellar luminosity L normalized to the solar luminosity $L_{sun}$, and the outer and inner edge distances (d) in AU can be determined using equation (1) (ibid):

$$d(AU) = \sqrt{\frac{L/L_{sun}}{S_{EFF}}} \qquad (1)$$

All our atmospheres are fully-saturated although a moist $CO_2$ adiabat is followed in the upper atmosphere. Six solar zenith angles are used in our calculations. We set the model surface albedo to 0.31, which reproduces the mean surface temperature for Earth (288 K) in our model. As with previous work on the HZ (Kasting et al., 1993; Kopparapu et al 2013, Ramirez and Kaltenegger, 2017), this value is higher than Earth's surface albedo (~0.2) and is designed to account for the additional reflectivity as well as heating effect from clouds. Cloud effects continue to be poorly understood, however, particularly for atmospheres very different from the Earth's. Thus, we do not vary cloud properties for our atmospheres (see e.g. Kasting et al 1993, Zsoms et al 2012, Kopparapu et al 2013, Kitzmann et al 2016). However, the planetary albedo is a sum of atmospheric and surface reflection, evolving as the atmospheric composition changes.

## 3. RESULTS

*Methane heats or cools a planet depending on the host star's SED*

The stellar radiation for 3 representative star types (F0, Sun, and M3) are shown in Figure 1. A significant portion of an M-star's stellar radiation is emitted at near-infrared wavelengths, particularly between 1 – 3 microns. Figure 2 shows 3 different atmospheres with $CO_2$, $CH_4$ and $H_2O$ concentrations representative of worlds near the outer edge of the habitable zone, where $CO_2$ dominates the atmospheric composition.

The near-infrared absorption cross-sections for $CH_4$ between ~1 and 4 microns are comparable to those in the IR near and around the 7.7micron band (Fig. 2) for the atmospheres shown/modeled. In contrast, the largest $CO_2$ and $H_2O$ infrared absorption cross-sections are near the Planck function peak and are generally orders of magnitude larger than their near-infrared ones between ~1 and 2.5 microns. Moreover, $H_2O$ vapor concentrations and its resulting absorption are relatively weak in dense $CO_2$ atmospheres near the outer edge (Fig. 2).

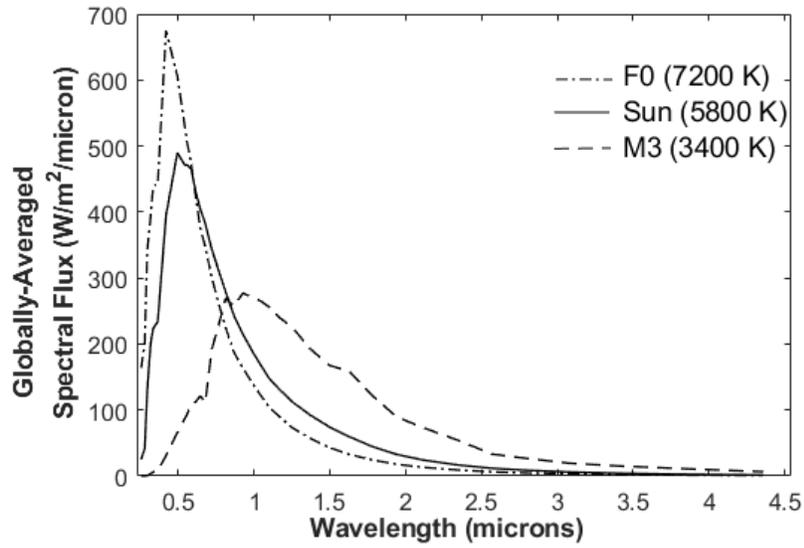

**Figure 1:** Globally-averaged incoming stellar spectra for the Sun, F0, and M3 stars.

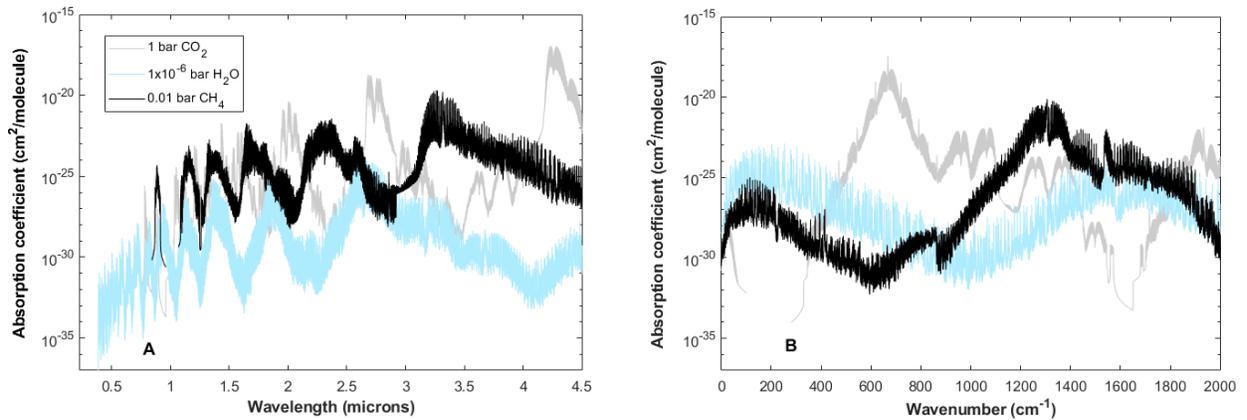

**Figure 2:** Illustration of absorption cross-sections for CH$_4$ (black), CO$_2$ (grey), and H$_2$O vapor (blue) as a function of (a) solar and (b) thermal infrared wavelengths (right) in a 1-bar atmosphere. The CO$_2$ case assumes a pure (fCO$_2$ = 1) CO$_2$ atmosphere whereas the background gas in the atmospheres containing CH$_4$ (fCH$_4$=1x10$^{-2}$) or H$_2$O (fH$_2$O = 1x10$^{-6}$) was assumed to be terrestrial air.

We model the effect of the addition of methane on the temperature profiles of $CO_2$-rich atmospheres near the outer edge of the habitable zone of host stars with $T_{eff}$ = 7200 K, 5800, K, and 3400 K (F0, Sun, and M3 spectral classes), respectively (Fig. 3). For ease of comparison, the same stellar insolation value ($S/S_o$ = 0.33) and dry atmospheric (minus water vapor) composition (1 bar $N_2$ and 3 bar $CO_2$) and a 4 bar surface pressure is shown in Fig. 3. We present results for (10,000 ppm) 1% $CH_4$ simulations to compare against those with no $CH_4$, which changes the surface pressure by about 0.4 bar between the two cases shown. Sensitivity studies (not shown) revealed the same general behavior to the addition of methane at both higher and lower $CH_4$ concentrations.

In the $CH_4$-free simulations, the surface temperature is greatest for the planet orbiting the M3 (3400 K) host star (279 K versus 201 K for the F0 planet) (dashed lines Fig. 3) because its SED is shifted towards longer wavelengths where Rayleigh scattering is reduced (given that Rayleigh scattering is proportional to $1/\lambda^4$), lowering the overall planetary albedo.

The addition of 1% $CH_4$ to the model planet (Fig. 3 solid lines) produces sufficient upper atmospheric heating to generate a temperature inversion in all three model planets. The surface temperature increases when 1% $CH_4$ is added for the planet orbiting the F0 host star (~18K) and the solar analog (~29K). However, the surface temperature decreases for the planet orbiting an M3 host star (~31K) because the red-shifted SED of its M3 star produces stronger upper atmospheric heating, which reduces the stellar energy available to heat the lower atmosphere and surface (Fig. 4). In contrast, modeled planets orbiting the F0 host star and the Sun show less upper atmospheric heating from near-IR absorption (Fig. 3) and smaller atmospheric temperature inversions. The upper atmospheric heating for planets orbiting a M3 host star cools their surfaces more efficiently than the increased Rayleigh scattering for planets orbiting an F0 star.

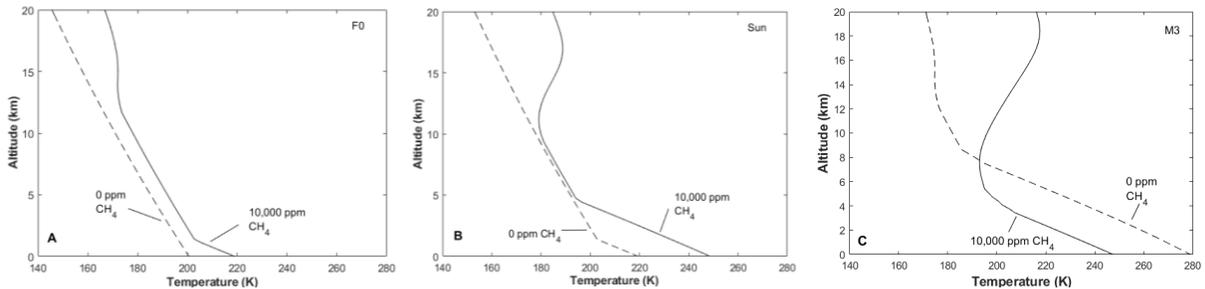

**Figure 3:** Changes in temperature profiles for a 3 bar $CO_2$ atmosphere ($S/S_o$ = 0.33) for a model planet with a surface pressure of about 4 bar orbiting a) an F0, b) a solar-analog, and c) an M3 host star when adding 1% (10,000 ppm) $CH_4$. A temperature inversion forms for all three model planets when $CH_4$ is added to the atmosphere. The surface temperature increases when 1% $CH_4$ is added for the planet orbiting the F0 (18K) and solar analog (29K), but decreases for the planet orbiting an M3 star (31K).

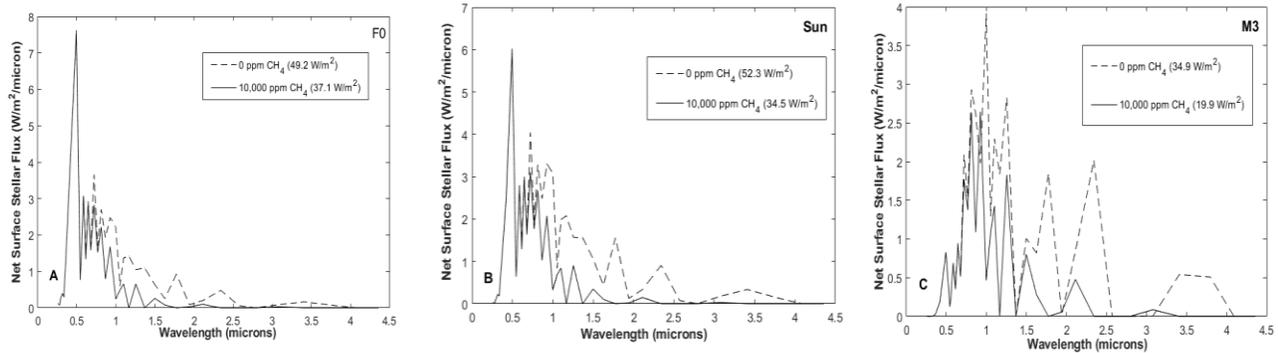

**Figure 4:** Changes in net surface spectral fluxes for a 4-bar total surface pressure, 3-bar $CO_2$ atmospheres around different host stars (temperature structures are shown in Figure 3). The addition of 1% (10,000 ppm) $CH_4$ reduces the NIR net stellar radiation received at the surface for the F0, Sun and M3 by 12.1, 17.8, and 15 W/m$^2$, respectively, which is equivalent to decreases of 25% (F0), 34% (Sun), and 43 % (M3).

*Methane expands the classical habitable zone outwards for hot host stars but reduces the HZ width for cool host stars*

The effect of adding different amounts of $CH_4$ to the classical $N_2$-$CO_2$-$H_2O$ habitable zone are shown in Figure 5 for the whole habitable zone and in Figure 6 for the outer edge. An alternative HZ limit that is not based on atmospheric models (like the classical HZ) but on empirical observations of our Solar System, is shown in both figures for comparison, the empirical HZ. The inner edge of the empirical HZ is defined by the stellar flux received by Venus when we can exclude the possibility that it had standing water on the surface (about 1Gyr ago), equivalent to a stellar flux of $S_{EFF}=1.77$ (Kasting et al.1993). The "early Mars" limit is based on observations suggesting that the martian surface may have supported standing bodies of water ~3.8 Gyr, when the Sun was only 75% as bright as today. For our solar system, $S_{EFF} = 0.32$ for this limit, corresponding to a distance of ~ 1.77 AU (e.g. Kasting et al. 1993). The incident flux in both figures is normalized to that received by Earth's orbit (~1360 W/m$^2$), $S_0$.

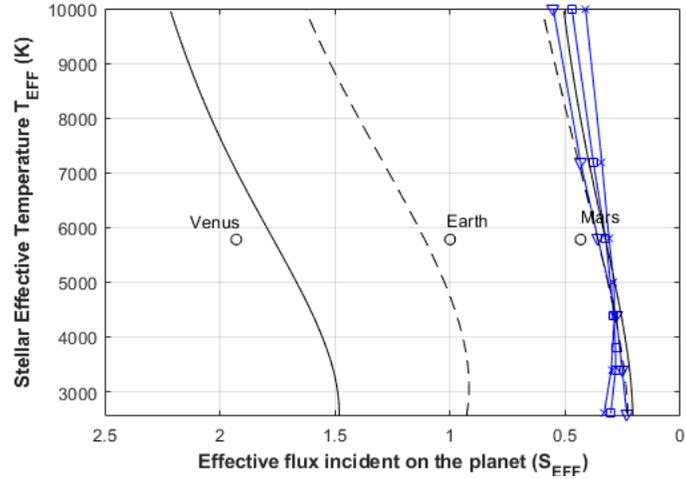

**Figure 5:** The effect of adding methane to the outer edge of the classical HZ. Stellar effective temperature versus incident stellar flux ($S_{EFF}$) for the traditional (dashed), empirical (solid black) and $CO_2$-$CH_4$ Habitable Zone (solid blue), respectively, for mixing ratios of 10 ppm $CH_4$ (triangle), 1% $CH_4$ (square), and $fCH_4 = 0.1 \times fCO_2$.

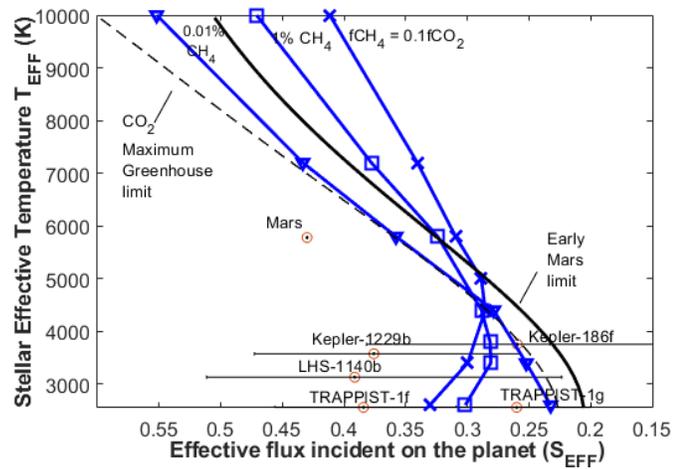

**Figure 6:** The effect of adding methane to the outer edge of the classical HZ. Stellar effective temperature versus the effective stellar flux ($S_{EFF}$) for the outer limits of the Habitable Zone. The traditional outer HZ limit, the $CO_2$ maximum greenhouse limit (dashed), is shown along with the empirical outer edge (solid black) and outer edge limits (solid blue) containing various amounts of $CH_4$: 10 ppm (triangle), 1% (square), and $fCH_4 = 0.1 \times fCO_2$. Some confirmed planets near the outer edge are included with error bars displayed.

Figures 5 and 6 show that $CH_4$ decreases the insolation needed at the outer edge of the habitable zone for hotter stars (~> 4500 K). Thus, the effect of $CH_4$ on the HZ depends on the host star's SED with lower stellar effective temperatures resulting in a $CH_4$ anti-greenhouse effect. As a haze is expected to build up for $CH_4/CO_2$ ratios greater than ~0.1 (Haqq-Misra et al. 2008), we do not explore even higher $CH_4$ mixing ratios.

The addition of $CH_4$ has a substantial effect on the limits of the outer edge of the Habitable Zone in a solar system like our own, decreasing $S_{EFF}$ from ~0.357 at the classical maximum greenhouse limit to ~0.305 for $fCH_4 = 0.1 \times fCO_2$ (~14.5%) (Figs. 5 - 6), equivalent to moving the outer edge from ~1.67 AU to 1.81 AU, beyond the empirical limit of the HZ. This lies just beyond the early Mars limit and is consistent with recent calculations suggesting that $CO_2$-$CH_4$ collision-induced absorption could have been significant on early Mars if $CH_4$ concentrations were sufficiently high (Ramirez, 2017; Wordsworth et al. 2017).

For planetary systems around hotter stars than the Sun, the addition of ~10ppm $CH_4$ decreases the $S_{EFF}$ required to support the outer edge for the hottest star in our sample, an A-star ($T_{eff}$ = 10,000 K) by ~8% from 0.598 to 0.552. Assuming $L/L_{sun}$ = 20, the orbital distance of the outer edge of the habitable zone moves from 5.78 to 6.01 AU. For $fCH_4 = 0.1 fCO_2$, $S_{EFF}$ decreases by as much as 31% (0.598 to 0.412) for the hottest sample star, moving the orbital distance of the outer edge of the habitable zone outward by over 1 AU (~21%), thus increasing the HZ width by 1.2 AU.

For host stars cooler than about 5000 K the addition of $CH_4$ leads to a net *cooling*. For the coldest star in our sample, an M8 star model at 2600 K, $S_{EFF}$ *increases* by ~3% for a concentration of ~10ppm $CH_4$. Assuming $L/L_{sun}$ ~ 0.0006, the orbital distance of the outer edge of the habitable zone moves from 0.0519 to 0.0512 AU. For $CH_4$ concentrations of ~1%, $S_{EFF}$ increases by as much as 46 % (0.2265 to 0.3305) (Figs. 5- 6) for the coolest sample star, moving the orbital distance of the outer edge of the habitable zone inward by over 0.009 AU (~18%). As with the greenhouse effect in the hotter stars, this anti-greenhouse effect becomes more pronounced in cooler stars as $CH_4$ concentrations increase (Figs. 5 - 6).

The addition of $CH_4$ reduces the $CO_2$ pressures required to achieve warm conditions. Whereas 8 bars of $CO_2$ are required at the outer edge of the solar system's classical HZ (e.g. Kasting et al. 1993; Kopparapu et al. 2013), this decreases to only 5 bar here for the high $CH_4$ case. The corresponding $CO_2$ pressures for A (10.000 K), F (7,200), and K-stars (5,000 K) are 4.5, 5, and 6 bar, respectively. In comparison, the equivalent $CO_2$ pressures for the classical outer edge are ~5.5, 6, and 8 bar.

Our calculations here replace the original $CO_2$-$CH_4$ HZ calculation in the Ph.D. thesis of Ramirez (2014), which had used an older $CO_2$-$CH_4$ parameterization. Overall, the general trends in both sets of results are similar to one another although the outer edge distance moves farther out for hotter stars in these new results.

Following previous studies (e.g. Kopparapu et al., 2013; Ramirez and Kaltenegger, 2017) we express a fourth order polynomial curve fit for this methane HZ expansion ($N_2$-$H_2O$-$CO_2$-$CH_4$ atmospheric composition):

$$S_{EFF} = S_{sun} + a \cdot T^* + b \cdot T^{*2} + c \cdot T^{*3} + d \cdot T^{*4} \quad (2)$$

Where $T^* = (T_{eff} - 5780)$ and $S_{sun}$ is the $S_{EFF}$ value in our solar system. The quantities (a,b,c,d) are constants. We have tabulated the parameterization data for both the outer edge of the $N_2$-$H_2O$-$CO_2$-$CH_4$ HZ (Table 1) as well as for the classical and empirical HZ (Table 2). Our parameterizations are valid for a larger spectral range than that in previous work (e.g. Kopparapu et al., 2013), expanded to include A spectral classes ($T_{EFF}$ = 2,600 – 10,000 K).

**Table 1.** Coefficients to calculate the outer edge of the Methane –HZ expansion ($N_2$- $CO_2$-$H_2O$-$CH_4$) for host stars with $T_{EFF}$ = 2,600 – 10,000 K

| Constant | 10 ppm $CH_4$ | 1% $CH_4$ | 0.1 $CH_4/CO_2$ |
|---|---|---|---|
| $S_{EFF(Sun)}$ | 0.3541 | 0.3248 | 0.3050 |
| A | 5.5923x10$^{-05}$ | 3.4781x10$^{-05}$ | 2.2160x10$^{-5}$ |
| B | 1.8197x10$^{-09}$ | 3.2218x10$^{-09}$ | 4.1913x10$^{-09}$ |
| C | -1.0257x10$^{-12}$ | -1.2967x10$^{-12}$ | -1.3177x10$^{-12}$ |
| D | 2.0424x10$^{-17}$ | 1.2453x10$^{-16}$ | 1.1796x10$^{-16}$ |

**Table 2.** Revised coefficients for the classical HZ (atmospheric composition $N_2$-$CO_2$-$H_2O$) and the empirical HZ for host stars with $T_{EFF}$ = 2,600 – 10,000 K

| Constant | Recent Venus | Leconte et al. | $CO_2$ Maximum Greenhouse | Early Mars |
|---|---|---|---|---|
| $S_{EFF(Sun)}$ | 1.768 | 1.105 | 0.3587 | 0.3246 |
| A | 1.3151x10$^{-04}$ | 1.1921x10$^{-04}$ | 5.8087x10$^{-05}$ | 5.213x10$^{-5}$ |
| B | 5.8695x10$^{-10}$ | 9.5932x10$^{-09}$ | 1.5393x10$^{-09}$ | 4.5245x10$^{-10}$ |
| C | -2.8895x10$^{-12}$ | -2.6189x10$^{-12}$ | -8.3547x10$^{-13}$ | 1.0223x10$^{-12}$ |
| D | 3.2174x10$^{-16}$ | 1.3710x10$^{-16}$ | 1.0319x10$^{-16}$ | 9.6376x10$^{-17}$ |

## 4. DISCUSSION

*Inverse calculations overestimate the effect of Methane heating – a sensitivity study*

As effective as $CH_4$ is in increasing the width of the habitable zone for hotter stars (Figs. 5 - 6), competing absorption at solar wavelengths reduces some of its effectiveness (Pierrehumbert, 2010). We show the results derived from inverse calculations to assess what the effect on the $N_2$-$H_2O$-$CO_2$-$CH_4$ outer edge would be if a constant stratospheric temperature (155 K) were assumed (equivalent to the assumption that upper atmospheric absorption was inefficient) (Fig. 7). Such a scenario would overestimate the effect of $CH_4$ absorption, reducing the $S_{EFF}$ in our solar system to 0.292 and 0.246, for 1% and ~0.1 $CH_4/CO_2$, respectively. This would correspond to overestimated outer edge distances of 1.86 and 2.02 AU, respectively. These are increases of ~0.13 AU and 0.21 AU (~8% and 12 % error), accordingly, over the actual values computed from forward calculations. These calculations suggest that other gases that can significantly heat the upper atmosphere, like ozone, cannot be accurately modeled via inverse calculations. To improve the reliability of results, forward calculations should be used for atmospheres that contain such atmospheric constituents.

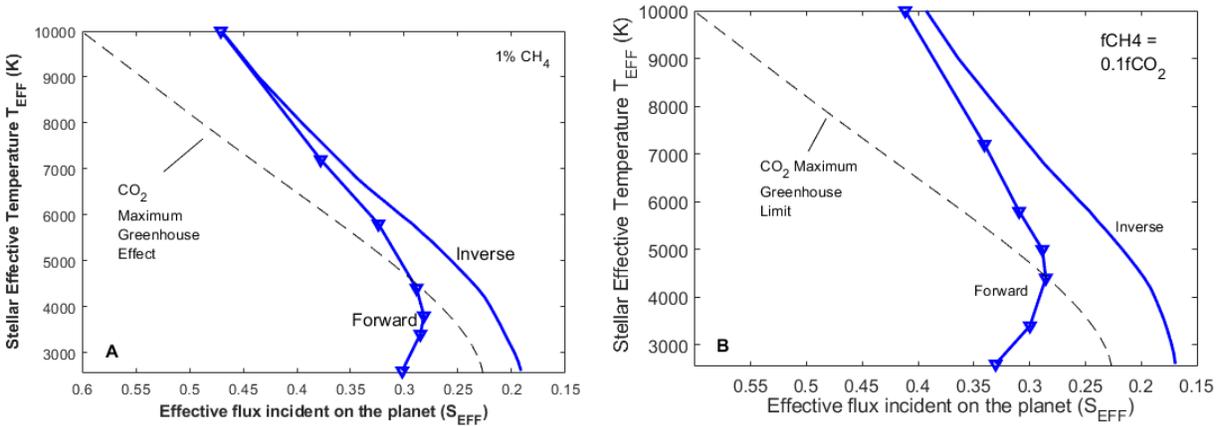

**Figure 7:** Comparisons of the modeled stellar effective fluxes ($S_{EFF}$) at the outer edge of the $N_2$-$H_2O$-$CO_2$-$CH_4$ Habitable Zone when using forward calculations for a) 1% (solid blue with triangles) and b) 0.1 $CO_2$ (solid blue with triangles) $CH_4$ cases versus the equivalent inverse calculations (solid blue line) for main-sequence (A – M) stars. The discrepancy between the forward and inverse computations is smallest for the hottest stars ($T_{eff} >\sim$ 9000K) because upper atmospheric absorption for methane on such planets is very weak. The error increases with decreasing stellar effective temperature of the host star.

*A-star habitability*

Our recent papers (Ramirez and Kaltenegger, 2016;2017) have expanded the stellar effective temperature range of stars relevant for the habitable zone from ~7,200 to 10,000 K. This reflects the notion that life on Earth may have evolved as quickly as ~700 million years (Mojzsis et al., 1997), which suggests that mid-late A-stars could be possible hosts for life (Danchi and Lopez, 2013; Ramirez and Kaltenegger, 2016).

*M-star outer edge HZ's dependence on Methane concentration*

The width of the HZ narrows for M-star planets near the outer edge as atmospheric $CH_4$ concentrations increase (Figs. 5 -6). Therefore, planets with $CH_4$-rich atmospheres near the outer edge of the classic HZ of M-stars should have frozen surfaces, according to this study. Such conditions may not bode well for the habitability of these worlds. Nevertheless, our results predict a $CH_4$ partial pressure gradient towards the outer edge of M-stars that can be tested by future observations observing atmospheres rich in both $CO_2$ and $CH_4$.

*The effect of $CO_2$-$CH_4$ Hazes*

Our models assume $CH_4$ concentrations no higher than ~10% of the $CO_2$ concentration, above which cooling photochemical hazes form as discussed earlier. Even though the inner edge would be relatively robust to increases in $CH_4$ at the concentrations considered, at even higher concentrations, the cooling effect from these hazes may be effective in moving the inner edge inward (e.g. Arney et al. 2016). Future work could assess this possibility, studying the effects of hazes composed of different particle sizes and shapes, including fractals (e.g. Wolf and Toon, 2010) in addition to coupling photochemical and climate modeling calculations. Nevertheless, any results would greatly depend on the assumed properties of the haze.

Our results here are unaffected even if hazes do not cool M-star planets (Arney et al. 2017). This is because the anti-greenhouse effect we find here for such stars is not due to haze and would occur anyway. Given the hypothesis that stars with high FUV fluxes may not produce haze (Arney et al. 2017), we performed a sensitivity study at a high $CH_4$ concentration (20%) for our A-star planet. The outer edge moves from 5.78 AU to 7.18 AU in this case (~24% increase), as compared to a 21% increase for the $fCH_4 = 0.1fCO_2$ case (see Results).

*Estimates of the occurrence rate of other Earths "$n_{earth}$"*

The atmospheres considered here, suggest that the outer edges of hotter (~A – G spectral classes) stars may extend proportionately farther out than those around cooler (K - M) stars at a given $CH_4$ concentration. Our models show that the classical habitable zones around cooler stars can *shrink* if planetary atmospheres contain even small amounts (ppm levels) of $CH_4$. This suggests that estimates of "$n_{earth}$", which describes the fraction of terrestrial planets within the liquid water habitable zone, is both sensitive to the composition of the atmosphere of the planet as well as the host star's SED.

*The plausibility of high $CO_2$-$CH_4$ atmospheres and the importance of life*

To assess the plausibility of our $CO_2$-$CH_4$ atmospheres we evaluate the various atmospheric sources (volcanism, serpentinization) and sinks (photolysis and $H_2$ escape) for $CH_4$ on a hypothetical Earth-sized planet with a 5.5 bar $CO_2$-$CH_4$ atmosphere (5 bar $CO_2$, 9% $CH_4$; neglecting the $N_2$) located near the outer edge (1.8 AU) of a G2-star.

We use the following weathering rate parameterization from Berner and Kothvala (2001) to estimate volcanic outgassing rates, assuming that weathering and volcanic outgassing rates are equal at steady state:

$$\frac{W}{W_{EARTH}} = \left(\frac{pCO_2}{p_{EARTH}}\right)^{\beta} e^{[k_{act}(T_{surf}-288)]} \left[1+k_{run}(T_{surf}-288)\right]^{0.65} \quad (3)$$

Here, W is the weathering rate, $k_{act}$ is an activation energy, $T_{surf}$ is surface temperature, $k_{run}$ is a runoff efficiency factor, β is the dependence of $pCO_2$ on W. $W_{earth}$ and $P_{earth}$ are the soil weathering rates and soil $pCO_2$ values, respectively, for the Earth.

This parameterization assumes that weathering/volcanic outgassing rates scales with pressure, as predicted for planets that have an operational carbonate-silicate cycle near the outer edge (e.g. Bean et al. 2017). Following Ramirez (2017), we assume β = 0.4, which is consistent with experimental measurements for silicate rocks (Lasaga, 1984; Asolekar et al, 1991; Schwartzman and Volk et al. 1989). We assume that soil $pCO_2$ is ~30 times that of atmospheric $pCO_2$, which assumes that life, particularly vascular planets, are present (Batalha et al. 2016). Although such high $CO_2$-$CH_4$ atmospheres are likely to be photochemically unstable as $CH_4$ slowly converts into CO (e.g Zahnle, 1986; Karecha et al. 2005), if such planets are inhabited, CO would have likely been consumed by microorganisms and converted back to $CH_4$ and $CO_2$ (e.g. Karecha et al. 2005).If we assume a relatively low mean surface temperature of 275 K, the $CO_2$ outgassing rate to support this atmosphere is ~$2.72 \times 10^{15}$ g/yr, which is ~6.9 times that computed for the Earth ($3.3 \times 10^{14}$ g/yr; Holland, 1984). The corresponding $CH_4$ production rate is ~$2.5 \times 10^{14}$ g/yr, which may be produced abiotically and/or biotically (see below). Assuming no atmospheric mixing, these numbers are equivalent to $CO_2$ and $CH_4$ global production rates of $2.5 \times 10^{11}$ and $5.8 \times 10^{10}$ molecules/cm$^2$/sec, respectively. If we neglect sinks, such production rates can produce 1 bar of $CO_2$ and 0.1 bar of $CH_4$ in ~2 million years. This volcanic $CO_2$ would be further supplemented by that retained from the primordial inventory. Earth, for instance, may have accumulated ~ 60 bars equivalent from accretion (e.g. Walker, 1981). Hydrodynamic escape rates during the early days of the stellar system for a planet located at 1.8 AU would be ~ 1/3 that at 1 AU, suggesting larger $CO_2$ inventories than Earth's may be possible for our planets. Moreover, hydrodynamic escape rates would be lowest for outer edge planets orbiting F- and A-stars because of their increased orbital distance and lower pre-main-sequence stellar luminosities as compared to their main-sequence values (e.g. Ramirez and Kaltenegger, 2014). Nevertheless, such high $CO_2$ outgassing rates would require an oxidized mantle, which is predicted for larger terrestrial planets, like Earth, shortly after they form (e.g. Wade and Wood, 2005; Hamano et al. 2013). Although this may suggest low $CH_4$ volcanic outgassing rates (e.g. Kasting and Catling, 2003), CH$_4$ may be produced through a few other ways.

CH$_4$ may be generated via serpentinization, which is a process by which Fe-rich waters produce H$_2$ via oxidation of basaltic crust (e.g. Chassefierre et al. 2013). On Earth, the methane flux produced by this process can be bounded by comparing it with the rate at which seafloor is oxidized by this process, resulting in the following equation (Shaw, 2014):

$$CO_2 + 2H_2O \leftrightarrow CH_4 + 2O_2 \qquad (4)$$

Serpentinization of seafloor on Earth produces ~$2\times10^{11}$ moles/yr of O$_2$ (Sleep, 2005). If we assume that serpentinization produces only CH$_4$ (not H$_2$) an upper bound of $1\times10^{11}$ moles of CH$_4$ or $4\times10^{8}$ molecules/cm$^2$/sec are produced on the Earth according to the above equation. However, CH$_4$ production rates from serpentinization can be potentially much higher if such rocks have an ultra-mafic composition (e.g. Batalha et al. 2015). For example, if serpentinization occurs on present Mars, Etiope et al. (2013) estimated that CH$_4$ production rates within localized regions could be ~ $4.4\times10^{12} - 4.4\times10^{13}$ molecules/cm$^2$/sec. At such production rates, serpentinization would only need to occur on ~0.13 – 1.3% of the planet's surface area to produce the above-mentioned CH$_4$ production rates.

Other potential abiotic sources for CH$_4$ include impacts, although resultant atmospheric chemistries are highly-dependent on impactor composition (e.g. Schaefer and Fegley, 2010). Moreover, impacts could add or remove atmospheric mass depending on impactor and planetary properties (e.g. Melosh and Vickery, 1989). M-star HZ planets may be particularly susceptible to atmospheric erosion because tightly-packed orbits produce more energetic impactors (Lissauer, 2007), suggesting that HZ planets orbiting hotter stars may be able to accumulate atmospheric mass more efficiently in this manner. CH$_4$ may also be produced through hydrothermal activity (e.g. Shaw 2008).

In contrast, the major CH$_4$ sinks are chemical destruction via photolysis and hydrogen escape to space. If we assume that escape is limited by its diffusion through the homopause, which is the fastest possible escape rate at these concentrations, then:

$$\phi = \frac{b}{H_a} fH_2 \qquad (5)$$

Here, fH$_2$ is the total H$_2$ mixing ratio at the homopause, H$_a$ is the atmospheric scale height, and b is a constant that describes diffusion of H$_2$ in a CO$_2$-dominated atmosphere (Zahnle et al. 1988). We assume a H$_2$ concentration of ½%. Given high CO$_2$ cooling rates (Wordsworth and Pierrehumbert, 2013), homopause temperatures for CO$_2$-dominated atmospheres should be low (300 – 500 K), suggesting nominal H$_2$ escape rates of ~5.5 – $8.5\times10^{10}$ molecules/cm$^2$/sec. This calculation assumes that H$_2$ escape is dominated by CH$_4$ produced by the presence of life, increasing the escape rate by a factor of 2 (e.g. Pavlov et al. 2000). However, the diffusion-limited escape rate may overestimate the actual one because if the exobase is cold, as expected in a low O$_2$ high CO$_2$ atmosphere, escape rates may follow the slower energy limit (e.g. Pavlov et al. 2001). Moreover, spherical geometry effects could decrease escape rates by a factor of 4 (Stone and Proga, 2009). Magnetic fields may further reduce H$_2$ losses (e.g. Stone and Proga, 2009 ; Ramirez et al. 2014) although we ignore such effects here. We only apply the geometric correction, calculating final $H_2$ escape rates of $\sim 1.4 - 2.1 \times 10^{10}$ molecules/cm$^2$/sec.

The second major loss process for atmospheres rich in methane is photolysis (e.g. Zahnle, 1986). An estimate for the maximum photodissociation rate of $CH_4$ for a planet located at Mars' distance from the present day Sun is $\sim 5 - 9 \times 10^{10}$ molecules/cm$^2$/sec (Wordsworth et al. 2017). The maximum $CH_4$ photodissociation rate at 1.8 AU should then be $\sim 3.6 - 6.4 \times 10^{10}$ molecules/cm$^2$/sec. Assuming, no other sources or sinks, the $CH_4$ would be completely removed in ~6 to 12 million years. However, these photolysis rates neglect reformation of dissociated $CH_4$ and absorption of escaping $H_2$, both of which would reduce $CH_4$ destruction rates below these numbers. Also, at the high $CH_4$ concentrations considered here, photolysis is driven primarily by Lyman-alpha photons (e.g. Zahnle, 1986; Pavlov et al. 2001). The latter suggests that photolysis rates should be considerably lower for planets orbiting stars hotter than the Sun, not only because of increased semi-major axis distances, but because Lyman-alpha emission seems to generally decrease with an increase in stellar effective temperature (e.g. Bohm-Vitense and Woods, 1983; Landsman and Simon, 1993). For example, an outer edge HZ planet located at 3.5 AU orbiting an F0 star would only need Lyman-alpha emission strength to be $\sim < 40\%$ that of the Sun for the aforementioned photolysis rates to decrease by an order of magnitude.

Overall, the total loss rates (photolysis plus $H_2$ escape) for $CH_4$ are of the same order as that of our computed volcanic outgassing rates for our solar case (although such loss rates may be lower for planets orbiting hotter stars). In the absence of life, these global production rates would decrease by a factor of $(30^\beta) \sim 4$, although this would be partially offset by lower $H_2$ escape rates in that scenario. Moreover, serpentinization could potentially complement volcanism by increasing $CH_4$ yields by a comparable (if not greater) amount. Moreover, simply increasing the surface temperature of our hypothetical planet by 10 K (to 285 K), would increase $CO_2$ and $CH_4$ outgassing rates by nearly a factor of 4, possibly bringing the atmosphere back into balance. However, all of these factors ignore perhaps the most important one: life. On Earth, well over 90% of the $CH_4$ produced has a biological origin. The enormous biological input of methanogens would greatly increase $CH_4$ production rates. As discussed, $CO_2$ outgassing rates may not need to be so high either, as microbes can convert CO into $CO_2$ (and $CH_4$) as well. Nevertheless, although our atmospheres can be maintained for some amount of time without life, especially around stars even hotter than the Sun, such planets still need relatively large atmospheric sources of $CH_4$ to compensate for the significant losses. Thus, finding relatively dense $CO_2$-$CH_4$ atmospheres near the outer edge of the habitable zone of hotter stars could potentially signal inhabitance. Indeed, a new study found this link between $CO_2$-$CH_4$ atmospheres and inhabitance for the early Earth (Krissansen-Totton et al. 2018), as we have discovered here. If we find a $CO_2$-$CH_4$ rich world near the outer edge of the HZ of a hot star, it should be further explored.

*Daisyworld: Methane can stabilize the climate of planets near the outer edge of the HZ*

We propose a stabilizing feedback loop suggested for the Archean Earth (Domagal-Goldman et al. 2008), which may also occur for habitable planets near the outer edges of hotter stars. Our scenario is essentially the "Daisy World hypothesis" (Watson and Lovelock, 1983) although the axes of the plots are reversed, with surface temperature exhibiting a parabolic response to increases in atmospheric $CH_4/CO_2$ ratio whereas $CH_4/CO_2$ exhibits linear responses to increases in surface temperature (Fig. 8). At low $CH_4/CO_2$ ratios, increases in $CH_4$ lead to increases in surface temperature because of the greenhouse effect of $CH_4$. This positive feedback explains why points on the left half of the curve (e.g. $P_1$) are unstable. Such a positive feedback may operate if methanogens could evolve on these planets, assuming that methane productivity increases with temperature (Domagal-Goldman et al. 2008). The resultant methanogenesis yields the following reaction, assuming some atmospheric $H_2$ is available:

$$4H_2 + CO_2 \rightarrow CH_4 + 2H_2O \qquad (6)$$

However, once $CH_4/CO_2$ ratios exceed ~ 0.1, photochemical hazes form and the surface cools until a stable warm point ($P_2$) just past the apex is reached. This point is stable because the slope of the line in the right half of the curve is negative. Further increases in surface temperature would lead to an increase in the $CH_4/CO_2$ ratio, thickening the haze and offsetting the warming with cooling (and returning the point leftward).

In contrast, the stabilizing feedback described would not operate on planets orbiting stars cooler than about 4500 K. This is because methanogenesis would lead to further increases in atmospheric $CH_4$, producing an *anti-greenhouse* effect on the planets orbiting these cooler stars, which would trigger a positive feedback and cooling until conditions could become too cold for surface life to exist (assuming life had arisen in the first place).

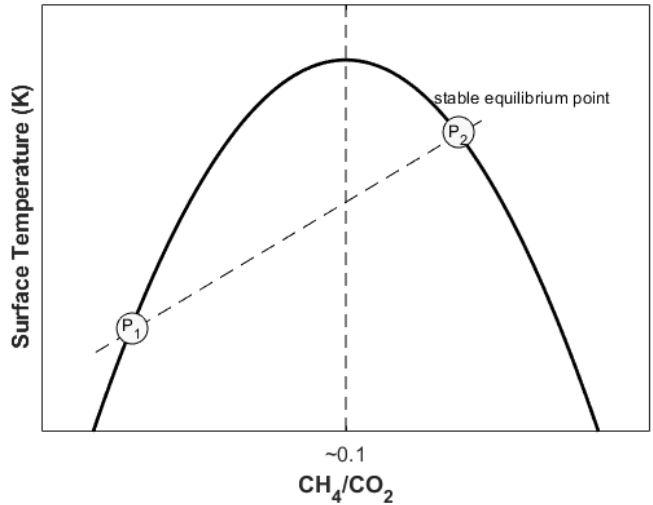

**Figure 8:** Proposed "Daisyworld" scenario for planets at the outer edge of the methane HZ with $CH_4$ in their atmospheres orbiting hotter (~A – G class) stars. The curved lines illustrate the effect that the $CH_4/CO_2$ ratio has on surface temperature whereas the straight lines depict the effect that temperature has on the $CH_4/CO_2$ ratio (adapted from Domagal-Goldman et al. 2008).

*Comparison of classical and $CO_2$-$CH_4$ HZ*

As had been shown previously (Ramirez, 2014; Ramirez and Kaltenegger, 2017) and in this work, secondary greenhouse gases are important to consider in HZ calculations. The percent changes to the outer edge distance from the addition of $CH_4$ are quite substantial (~ -20 to > 20%).

Moreover, the addition of $CH_4$ can significantly reduce the $CO_2$ pressures required to achieve warm conditions as compared to the classical HZ (see Results), potentially relaxing the requirements that models need to satisfy to achieve warm mean surface temperatures. Our results here are consistent with previous ones that found that the addition of $H_2$ also reduces the $CO_2$ pressures required to support warm outer edge atmospheres (Ramirez and Kaltenegger, 2017). This implies that carbonate-silicate cycle predictions suggesting that $CO_2$ pressures on potentially habitable planets should increase towards the outer edge (Kasting et al. 1993; Bean et al. 2017) may be complicated by the contribution of secondary greenhouse gases. Significant non-linearities could be introduced that could thwart a straightforward relationship between $CO_2$ pressure and orbital distance.

Ramirez and Kaltenegger (2017) had shown that the empirical outer edge limit (i.e. early Mars) distance can be exceeded by modest increases in secondary greenhouse gas concentrations (Figs. 5 - 6). At 1.81 AU, our computed outer edge for our solar system is still 0.04 AU greater than the corresponding empirical early Mars limit distance (1.77 AU) (Figs. 5- 6). Moreover, much larger margins are calculated for A- and F-stars. For example, the early Mars limit would be exceeded by ~0.68 AU (~11 %) in our A-star case (see Results). We also point out that if hazes do not form around F- to A-stars (Arney et al. 2017) (see Methodology), then even higher $CH_4$ concentrations than what we consider here might be possible, which would extend the outer edge of our $CO_2$-$CH_4$ HZ for those stars even farther out than what we considered here. Plus, the early Mars limit distance can be exceeded by well over 40% using $H_2$ as a secondary greenhouse gas (Ramirez and Kaltenegger, 2017). All of this suggests that the optimistic outer edge limit for the classical HZ is still a lower bound on the outer edge distance. Thus, the continued exploration of different greenhouse gas combinations will show which scenarios can extend the HZ outer edge (Pierrehumbert and Gaidos, 2011; Ramirez, 2014; Ramirez and Kaltenegger, 2017; Wordsworth et al., 2017; Ramirez et al., 2018). This may also help solve the faint young sun problem for outer edge planets like Mars (e.g. Ramirez et al.; 2014; Ramirez, 2017).

*Effects of Clouds and Planetary Rotation Rates*

Although we cannot assess multi-dimensional effects, like clouds and rotation rate variations, self-consistently using only our 1-D climate model, we have recently assessed the effects of rotation rate on ocean worlds near the outer edge using a combination of our 1-D and latitudinally-dependent energy balance climate models (Ramirez and Levi, 2018, in press). We also note that such effects have also been assessed for the inner edge of the classical HZ using 3-D models (e.g. Kopparapu et al. 2017). If the results of such studies are any indication, rotation rates and clouds may also influence the extent of our $CO_2$-$CH_4$ HZ.

## CONCLUSION

We show that adding methane to a terrestrial planet's atmosphere heats or cools it depending on the host star's spectral energy distribution because upper atmospheric absorption competes with that from the greenhouse effect. We assess the greenhouse effect of $CH_4$ (10 - ~100,000 ppm) on the outer edge of the habitable zone ($N_2$-$CO_2$-$H_2O$-$CH_4$) for main-sequence host stars for stellar temperatures of 2,600 – 10,000 K (A3 to M8). Adding $CH_4$ to the classical habitable zone ($N_2$-$CO_2$-$H_2O$) produces net greenhouse warming for planets orbiting stars hotter than about 4,500K (~K3), whereas $CH_4$ absorption produces an anti-greenhouse effect for planets around cooler stars. We parameterize the outer edge limits of this methane HZ (atmospheric composition $N_2$-$CO_2$-$H_2O$-$CH_4$).

Methane expands the classical habitable zone outwards for host stars with effective temperatures above about 4,500K but reduces the HZ width for cooler host stars. We show $CH_4$ concentrations that are 10% that of $CO_2$ can increase the width of the habitable zone of the hottest stars in our model grid (10,000 K) by over 20%. In contrast, that same $CH_4$ concentration can shrink the habitable zone for the coolest stars in our model grid (2,600 K) by a similar percentage. The classical empirical outer edge limit is also exceeded for G – M planetary systems. Lower required $CO_2$ pressures also relax requirements that models need to satisfy to simulate warm conditions.

We find that dense $CO_2$-$CH_4$ atmospheres found near the outer edge of the habitable zone of hotter stars could suggest inhabitance although we cannot completely rule out that abiotic processes, such as high volcanic outgassing and/or serpentinization rates, cannot also produce similar atmospheres. Moreover, lower $H_2$ escape rates for outer edge planets and potentially lower photolysis rates for worlds orbiting stars hotter than the Sun, could favor $CH_4$ buildup, even in the absence of life.

We also propose a stabilizing feedback loop for inhabited planets on the outer edge of the habitable zone for planets orbiting host stars with effective temperatures above ~4,500K.

These results highlight the importance of including secondary greenhouse gases, complementing $CO_2$ and $H_2O$, in alternate definitions of the habitable zone in our search for habitable worlds.

## ACKNOWLEDGEMENTS

We thank James F. Kasting, Ray Pierrehumbert, Richard Freedman, and David Crisp for the helpful discussions. We also acknowledge support by the Simons Foundation (SCOL # 290357, Kaltenegger), Carl Sagan Institute at Cornell, and the Earth-Life Science Institute.